\documentclass[a4paper,fleqn,usenatbib]{mnras}
\usepackage{newtxtext,newtxmath}

\usepackage[T1]{fontenc}
\usepackage{ae,aecompl}

\usepackage[english]{babel}
\usepackage{amsmath,amsfonts,amssymb,textcomp}
\usepackage{amssymb}
\usepackage{pdflscape}
\usepackage{multirow}
\usepackage{varioref}
\usepackage{hyperref}
\usepackage{psfig}
\usepackage{graphicx,graphics,subfigure}
\usepackage[usenames,dvipsnames]{xcolor}
\hypersetup{backref,
colorlinks=true,citecolor=blue, urlcolor=magenta}

\labelformat{equation}{\textup{(#1)}}

\labelformat{Fig.}{\textup{(#1)}}

\graphicspath{{images/}}

\title[LoCuSS: Pre-processing in infalling galaxy groups]{LoCuSS: Pre-processing in galaxy groups falling into massive galaxy clusters at z=0.2}
\author[M. Bianconi et al.]{
M. Bianconi$^{1,}\thanks{mbianconi@star.sr.bham.ac.uk}$, G. P. Smith$^{1}$, C. P. Haines$^{2}$, S. L. McGee$^{1}$, A. Finoguenov$^{3,4}$,
\newauthor \hspace{0.12cm}E. Egami$^{5}$\\\\
$^{1}$School of Physics and Astronomy, University of Birmingham, Edgbaston, Birmingham, B15 2TT, UK\\
$^{2}$INAF - Osservatorio Astronomico di Brera, via Brera 28, 20122 Milano, Italy\\
$^{3}$Max-Planck-Institut f\"{u}r extraterrestrische Physik, Giessenbachstra{\ss}e, 85748 Garching, Germany\\
$^{4}$Department of Physics, University of Helsinki, Gustaf  H\"{a}llstr\"{o}min katu 2a, FI-0014 Helsinki, Finland\\
$^{5}$Steward Observatory, University of Arizona, 933 North Cherry Avenue, Tucson, AZ 85721, USA
}

\date{Accepted XXX. Received YYY; in original form ZZZ}

\pubyear{2017}

\begin{document}
\label{firstpage}
\pagerange{\pageref{firstpage}--\pageref{lastpage}}
\maketitle

\begin{abstract}
We report direct evidence of pre-processing of the galaxies residing in galaxy groups falling into galaxy clusters drawn from the Local Cluster Substructure Survey (LoCuSS).  34 groups have been identified via their X-ray emission in the infall regions of 23 massive ($\rm \langle M_{200}\rangle = 10^{15}\,M_{\odot}$) clusters at $0.15<z<0.3$. Highly complete spectroscopic coverage combined with 24 $\rm\mu$m imaging from Spitzer allows us to make a consistent and robust selection of cluster and group members including star forming galaxies down to a stellar mass limit of $\rm M_{\star} = 2\times10^{10}\,M_{\odot}$. The fraction $\rm f_{SF}$ of star forming galaxies in infalling groups is lower and with a flatter trend with respect to clustercentric radius when compared to the rest of the
cluster galaxy population. At $\rm R\approx1.3\,r_{200}$ the fraction of star forming galaxies in infalling groups is half that in the cluster galaxy population. This is direct evidence that star formation quenching is effective in galaxies already prior to them settling in the cluster potential, and that groups are  favourable locations for this process.\vspace{0.5cm}
\end{abstract}
\section{Introduction}

\begin{figure*}
 \centering
 \includegraphics[width=0.60\linewidth, keepaspectratio]{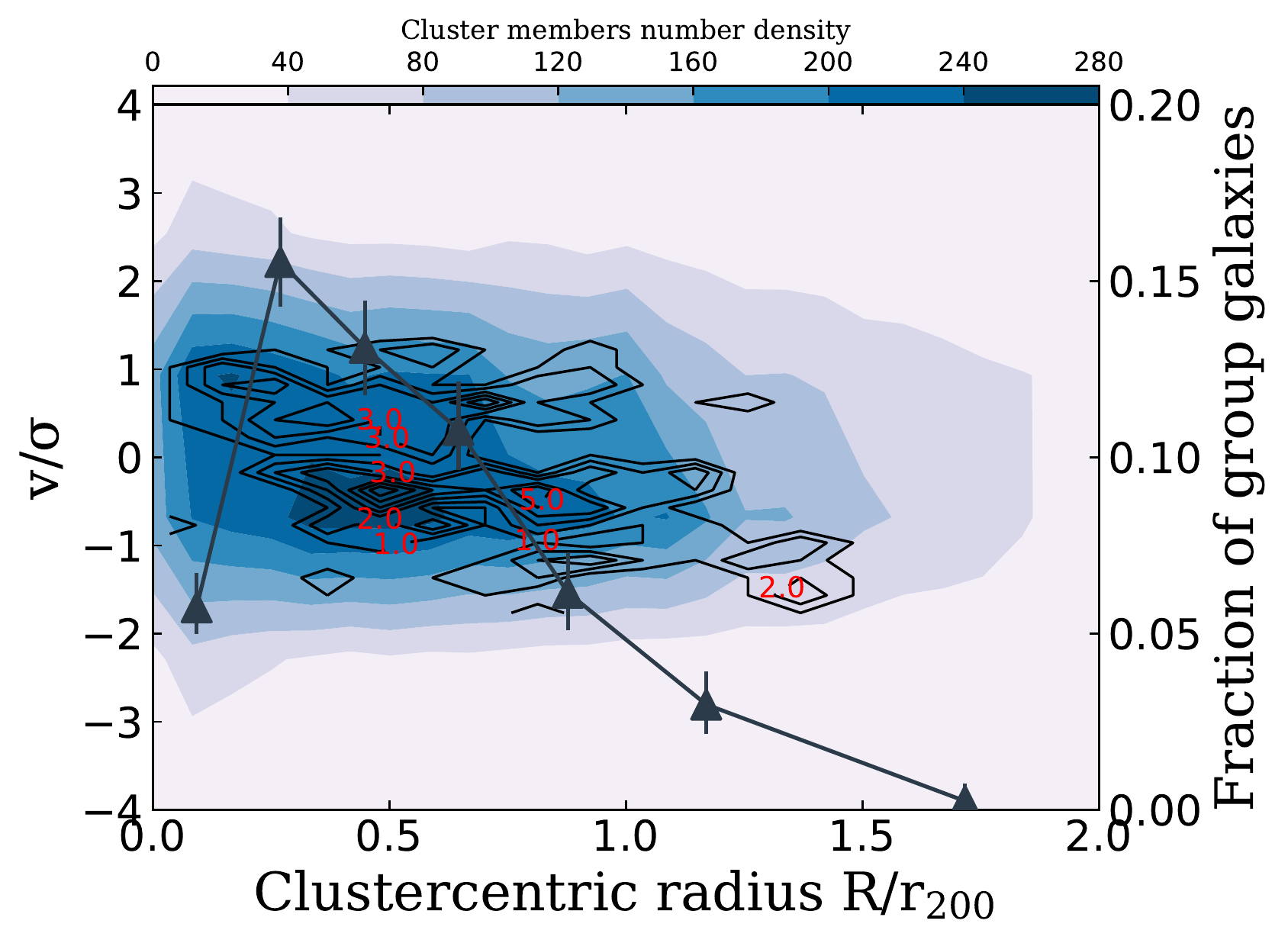}
\caption{Density map of the line-of-sight peculiar velocity (in units of $\rm\sigma$) versus clustercentric radius (in units of $\rm r_{200}$) of cluster members. Superimposed black contours are labelled in red with the number density of the 34 infalling group members. Filled gray triangles show the fraction of group members with respect to cluster members, as a function of clustercentric radius. The error bars (at $\rm 1 \sigma$) are computed from binomial statistics following \citet{gehrels86}.}\label{caustic}
 \end{figure*}

Massive galaxy clusters appear last in the hierarchical ladder of cosmic structure formation, according to current $\Lambda$CDM models and observations \citep{frenk12}. Cluster evolution is driven by accretion, in the form of both baryonic and non-baryonic matter, of smooth continuous streams of small haloes, together with episodic merging of more massive systems. As a result, clusters are located in dynamically active and overdense regions of the Universe. Observationally, dense environments found in clusters are also characterised by a dearth of spiral galaxies compared to the field (e.g. \citealt{dressler80}) and a star formation rate which shows a similar dependence on galaxy density (e.g \citealt{haines07, smith05}). While the accretion history must play a key role in shaping these galaxy populations, it is still unclear how. Indeed, the star formation transition between environments is not abrupt and the steady increase in the number of star forming galaxies goes together with the increase of distance from cluster centres (\citealt{wetzel12,haines15}). 
As shown in early studies \citep{couch98, dressler99}, the appearance of spiral galaxies with low star formation rates in the outskirts of clusters and the analysis of timescales of quenching processes, requires the presence of environmental effects accelerating the consumption of the gas reservoir prior to the settling of the galaxy into the cluster potential (so called pre-processing,  \citealt{zabludoff96, fujita04}). Further observations \citep{haines15, wetzel13} confirmed the deficit of star forming galaxies in the infalling regions of clusters, extending out to 5 virial radii, which can be successfully reproduced by simulations \citep{bahe13}.

In numerical simulations, massive clusters have accreted up to 50\% and 45\% of their stellar mass and galaxies, respectively, through galaxy groups \citep{mcgee09}. Hence, being fundamental building blocks of both mass and galaxy population in clusters, galaxy groups could play a significant role in shaping the evolution of cluster galaxies, if confirmed to be favourable locations for star formation quenching. However, no clear scenario is accepted regarding the nature and effects of pre-processing. For example, \citet{lopes14} argued that the main driver of galaxy evolution is local environment, both in groups and clusters. On the other hand\citet{hou14} suggested that pre-processing is the dominant cause of the quenched fraction of galaxies in massive clusters. Despite this, the role of group pre-processing, both in isolation and around clusters, remains unclear and requires further investigation (see also \citealt{lietzen12}).

Here we present a study of a new sample of recently discovered infalling groups \citep{haines17}, focussing in particular on the star formation properties of the massive galaxies within them. The extensive coverage of the dataset in both area and wavebands is coupled with a consistent infrared-based object selection, resulting in a unique sample of unambiguously selected infalling group galaxies which is ideal for the study of pre-processing.

This paper is structured the following manner. In Section~\ref{data}, we
present the datasets used. In Section~\ref{analysis}, we present the methods and main results of the data analysis. In Section ~\ref{summary}, we discuss the results and present future prospects of the project. Throughout this work, we assume $\rm H_0= 70\, km \,s^{-1} Mpc^{-1}$, $\rm\Omega_M= 0.3$ 
and $\rm\Omega_{\Lambda}= 0.7$.

\section{LoCuSS Groups}\label{data}
The galaxy groups considered here have been discovered in the infall regions of a subsample of clusters within the Local Cluster Substructure survey (LoCuSS). The LoCuSS survey targets  X-ray luminous clusters from the \textit{ROSAT} All Sky Survey catalogues (\citealt{ebeling98,ebeling00})  within a redshift range of $\rm 0.15\leq z \leq 0.30$. An extensive spectroscopic and photometric  campaign targeted the first 30 clusters that had  Subaru/Suprime-Cam ($V$ and $i'$ bands)  imaging. In particular, wide-field ($\rm \approx$1 degree diameter) optical spectroscopy with MMT/Hectospec was performed and composes the Arizona Cluster Redshift Survey (ACReS, \citealt{haines13}). Additionally, these clusters have been imaged in the near-infrared ($J$ and $K$ band) with UKIRT/WFCAM, covering $\rm 52'\times 52'$, mid-infrared 24$\rm \mu m$ with Spitzer/MIPS and far-infrared with Herschel/PACS and SPIRE, both covering $\rm 25'\times 25'$  (\citealt{haines10, smith10, pereira10}). Archival \textit{XMM}-Newton imaging is available for a subset of 23 out of these 30 clusters, extending out to $\rm \approx 2\,r_{200}$. These 23 clusters comprise the  sample for the detection of infalling groups and are listed in Table 1 of \citet{haines17}, together with the complete description of the observational campaign and data reduction strategy. Hereafter we introduce some salient aspects that we used in the current work.

Groups are identified from their X-ray extended emission.  The crucial step regards the separation of IGM extended emission from background and point sources. \citet{haines17} followed the technique of \citet{finoguenov09,finoguenov10,finoguenov15} consisting of signal decomposition via wavelets \citep{vikhlinin98} to detect signals above 4$\sigma$ on scales of 8 and 16 arcsec, which is associated to point sources and subtracted from large scales using the \textit{XMM} PSF model \citep{finoguenov09}.  Similarly, extended sources (e.g. groups) are associated to signals on scales of 32 and 64 arcsec above 4$\sigma$ from a 32 arcsec generated noise map. 

In order to confirm the extended X-ray detections as groups, we compared them with our optical imaging and spectroscopy. Sub-arcsecond resolution $V$ and $i$ band imaging is available from Subaru \citep{okabe10}. Spectroscopy is available from ACReS data. The ACReS survey was designed by selecting spectroscopic targets using the tight (0.1 mag) scatter of cluster galaxies in near infrared $J$-$K$ colour, following \citet{haines09a, haines09b}. As presented in \citet{haines13}, the colour cuts were tested by considering the $J$-$K$/$K$-band colour-magnitude diagrams of member galaxies for eight previously well-studied clusters, using only the literature redshifts (60-200 members per cluster). No a priori knowledge on the $J$-$K$ colour was used for the selection of these test galaxies. The $J$-$K$ color cut is found to be highly effective ($\rm98\%$) in retrieving confirmed spectroscopic members, irrespective of galaxy star-formation history or rate, as both star-forming and quiescent galaxies lie along the same sequence in $J$-$K$/$K$-band diagram. This allows for the selection of a stellar mass limited sample of galaxies, and for an effective removal of foreground stars with no additional a priori bias on galaxy spectroscopic type. The magnitude limit reached by the survey is $\rm M^*_K+1.5$, which corresponds to a mass limit of $\rm M_{*}\approx 2\times 10^{10} \, M_{\odot}$. In order to compensate for any selection bias of the spectroscopic targets due to intrinsic photometric properties and position on the plane of the sky, each galaxy is weighted by the inverse of the probability of being targeted by spectroscopy \citep{haines13}, as proposed by \citet{norberg02}. The spectroscopic completeness of the K-band magnitude limited sample is $\rm 80\%$ in the  \textit{XMM} covered areas. The spectroscopic completeness increases to $\approx96\%$ when considering $\rm 24 \mu$m-detected sources \citep{haines13}.
 
In order to associate galaxies to the intra-group medium (IGM) emission, a visual inspection of Subaru optical data was performed to firstly identify the central group galaxy as a massive, early-type galaxy close to the centre of X-ray emission. The galaxy companions are selected within $\rm 1000 \,km\,s^{-1}$ and 3 arcmin from the central group galaxy. In case of no clear central galaxy, a close ($\rm\approx 2 \, arcmin$ with $\rm \Delta v < 1000 \,km\,s^{-1}$) pair of galaxies is considered as the central object and the velocity and separation cuts are applied as before to select the other group galaxies. The analysis in \citet{haines17} shows that this strategy allows us to identify groups out to $\rm z \sim 0.7$. Furthermore, all X-ray group at $\rm z < 0.4$ (based on their photometric redshift) have at least one spectroscopically confirmed galaxy member.  

The result of  group identification and member selection compared to the LoCuSS cluster galaxies can be seen in Figure~\ref{caustic}. Here the number density of cluster and group galaxies is plotted in the clustercentric versus peculiar velocity phase space, highlighting also the fraction of members in groups with respect to cluster members as a function of clustercentric distance. A total of 34 groups, averaging $\approx$10 members each, is understood to be infalling onto clusters, due to their relative position and velocity with respect to their closer cluster. \citet{diaferio97} showed that the escape velocity of a galaxy cluster at a given radius can be traced by analysing the line-of-sight velocity distribution of galaxies with respect to their projected distance from the cluster centre (so-called "caustic" technique). In particular, galaxies within the characteristic structure of the caustic lines are gravitationally bound to the cluster, hence infalling or orbiting the cluster potential. Our groups are all located within the caustic lines of clusters (\citealt{haines17}, Figure~\ref{caustic}), in areas of the diagram associated to recent or ongoing infall on cluster haloes (\citealt{haines15,rhee17}). This suggest that the groups are at the first encounter with the cluster environment. Advanced stages of accretion are excluded since ram-pressure stripping would have removed the IGM from group haloes.

 \begin{figure*}
 \centering
 \includegraphics[width=0.65\linewidth, keepaspectratio]{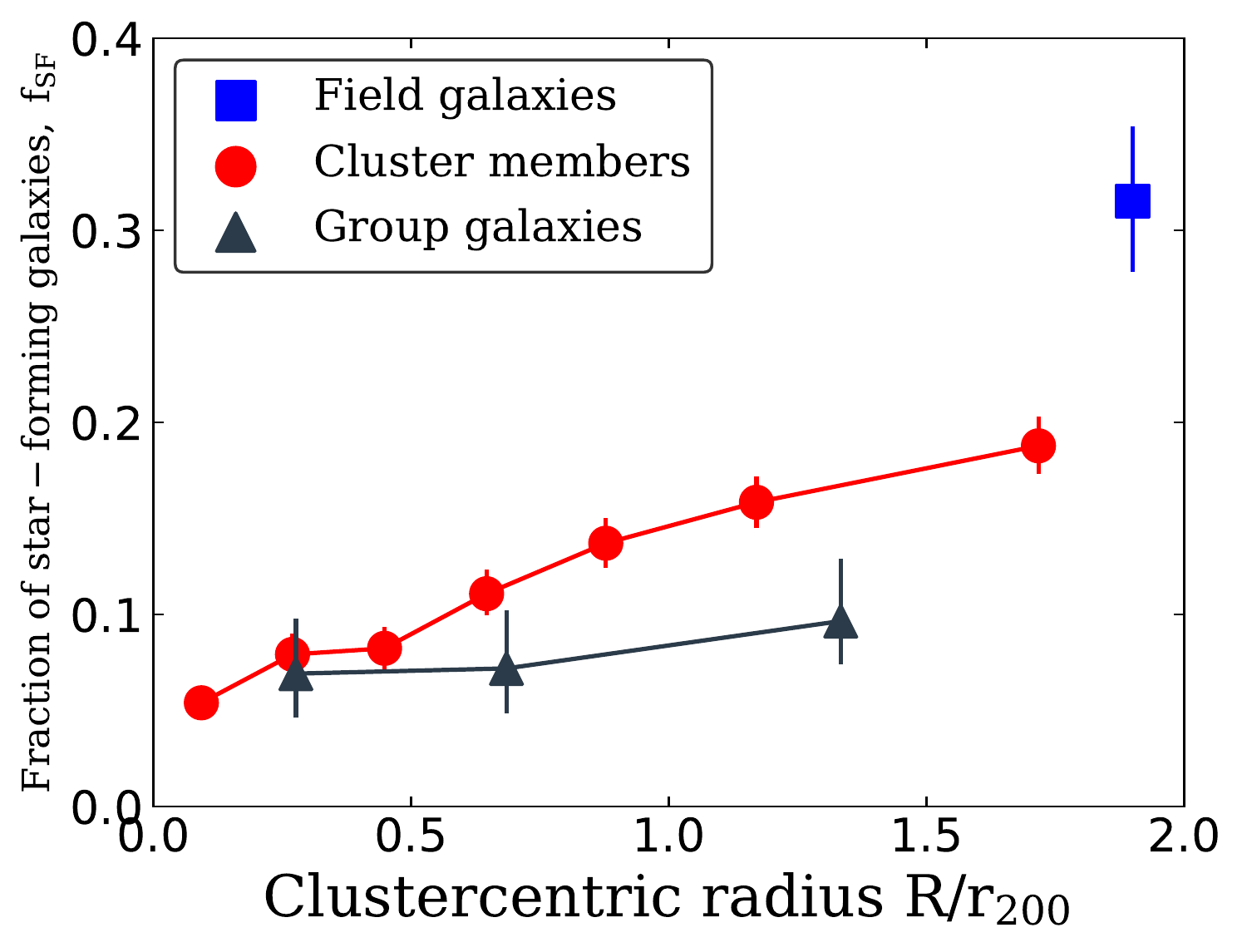}
\caption{Fraction of massive ($\rm M_* > 2\times 10^{10} \,M_{\odot}$) star forming($\rm SFR > 2\,M_{\odot}\,yr^{-1}$) galaxies, $\rm f_{sf}$, plotted with respect to projected clustercentric distance in units of $\rm r_{200}$. Galaxies in LoCuSS clusters and infalling groups are plotted as filled red circles and gray triangles, respectively. The $\rm f_{sf}$ of field galaxies is plotted as filled blue squares at an arbitrary radius.The error bars (at $\rm 1 \sigma$) are computed from binomial statistics following \citet{gehrels86}. Each radial bin contains on average 723 cluster and 125 group galaxies, respectively.}\label{sf_fraction}
 \end{figure*}

\section{Analysis and Results}\label{analysis}
In this work, we perform a careful group member selection by identifying galaxies with peculiar velocity  $\rm \left | v \right |= \left |c(z-\bar{z})/(1+\bar{z})\right |< 500 \,km\,s^{-1}$, computed with respect to each group mean redshift ($\rm \bar{z}$, following \citealt{harrison79}), and within each group $\rm r_{200}$.  Each group's X-ray luminosity $\rm L_{X}$ is used to obtain an estimate of $\rm M_{200}$ following \citet{leauthaud10}, and hence its $\rm r_{200}$. \citet{haines17} showed that these masses are compatible with estimates from each group velocity dispersion. Five of the groups selected in \citet{haines17} have been excluded from the current analysis for being part of highly disturbed systems (A115-8 and 10, A689-7 and 8) and A1758-8 for being having mass and size comparable to a cluster rather than a group. Stellar masses were computed using the linear relation between stellar mass-to-light ratio  and optical $g$-$i$ color from \citet{bell03},
\begin{equation}
\rm Log(M/L_K) = a_k + [b_k \times (g-i)]
\end{equation}
where $\rm a_k = -0.211 $,  $\rm b_k = 0.137 $. The relation is corrected by $\rm -0.15$ to be compatible with a \citet{kroupa02} initial mass function (see also \citealt{haines15}). The stellar mass of the group galaxies does not show any trend with redshift, with an average value of $\rm log M_{\star}[M_{\odot}]= 10.75$ across the considered redshift range. As shown by \citet{bell03}, the scatter in the relation ranges up to 0.1 dex, suggesting its robustness. The 24$\rm \mu$m luminosity of each source is obtained from a comparison between the observed flux and models of infrared spectral energy distributions by \citet{rieke09} and subsequently converted into obscured star formation rate following \citet{rieke09}
\begin{equation}
\rm SFR_{IR}\,[M_{\odot}yr^{-1}] = 7.8\times10^{-10}\,L_{24}\,[L_{\odot}]
\end{equation}
As quoted by \citet{rieke09}, the SFR evaluated from $\rm 24 \,\mu m$ emission is accurate within 0.2 dex when compared to SFRs obtained from a linear combination of total infrared luminosity and $\rm H_{\alpha}$ emission. Furthermore, $\rm L_{24}$  is a reliable proxy for normal galaxies in recovering the total infrared luminosity, hence allowing for a solid estimate of SFR (see also the review by \citealt{calzetti13}). Type-1 AGN can contribute to dust heating at $\rm\approx1000\,K$, dominating the emission of the host galaxy at 24 $\rm\mu$m \citep{xu15}. In order to avoid any contamination to our star formation estimates, we excluded AGN candidates selected via optical broad lines and X-ray emission. Additionally, $\rm98\%$ of our 24 $\rm\mu$m galaxies are also detected at 100 $\rm\mu$m with Herschel. At this wavelength, the infrared emission is dominated by star formation, excluding a significant  contribution from AGN. When considering clusters in the low reshift bin ($\rm 0.15 < z< 0.20$) $\rm 99\%$ of the sources above the $\rm SFR_{IR}$ threshold present  emission in $\rm H_{\alpha}$. This indicates that the infrared emission is indeed due to star formation and not evolved stellar populations \citep{haines15}. Given the depth of Spitzer observations, the lower limit of sensitivity for which star formation can be evaluated is $\rm SFR_{IR}\approx 2\,M_{\odot}yr^{-1}$. This value is also representative of the average $\rm SFR_{IR}\approx 5\,M_{\odot}yr^{-1}$ of the galaxies within the LoCuSS survey, in which only four objects are exceeding $\rm SFR_{IR}\approx 40\,M_{\odot}yr^{-1}$ \citep{haines13}.

The total number of selected infalling group members is 400 of which 375 satisfy our stellar mass cut ($\rm M_* > 2\times 10^{10} \,M_{\odot}$), and 29/375 satisfy  $\rm SFR > 2\,M_{\odot}\,yr^{-1}$ and are thus classified as star forming. Our main result is presented in Figure~\ref{sf_fraction}. The fraction of massive star forming galaxies  $\rm f_{sf}$ in clusters (red points) and the infalling groups (gray triangles) is plotted with respect to cluster-centric distance, and compared to that seen in coeval field population (blue square). We can see a clear decrease of massive star forming galaxies in clusters with respect to decreasing projected clustercentric distance. Both groups and clusters depart from the field population. Noticeably, the fraction of massive star forming cluster galaxies shows a steeper increasing trend with clustercentric radius compared to group galaxies. This culminates at around $\rm 1.3\,r_{200}$, at which the fraction of star forming group galaxies is about half of the fraction in clusters, with a significance of $\rm \approx 2\sigma$. The fraction of star forming galaxies does not depend on stellar mass within the mass range that we cover in our sample. We tested the sensitivity of our result to the star formation threshold that we adopt by gradually increasing it, resulting only in a continuous decrease of $\rm f_{sf}$ in field, clusters and groups. Furthermore, we checked the effect of contamination by cluster galaxies to group members by relaxing the membership cuts in radius and velocity, obtaining an increase of the fraction of star-forming galaxies in groups. The flat trend of $\rm f_{SF}$ in groups suggests that star formation quenching is effective in groups prior to their arrival in clusters. We consider this as direct evidence of pre-processing.   As groups are accreted, they  populate clusters with galaxies similar to those found in cluster cores. We tested the dependence of $\rm f_{SF}$ in groups with respect to distance from the group center, by removing galaxies at progressively smaller radii from the group X-ray peak. We did not find significant changes in the overall trend of $\rm f_{SF}$, other than an increase of the scatter.

\section{Discussion}\label{summary}

 In this study, we present direct evidence of group galaxies undergoing pre-processing, resulting in a fraction of star forming galaxies that is lower than cluster galaxies at the same cluster-centric distance. In clusters, the increase of the fraction of star-forming galaxies with cluster centric distance is due to relative increase of the number of star-forming galaxies in the infall regions with respect to passive galaxies, which dominate cluster cores \citep{haines15}. In this work, we separate galaxies infalling within groups from the remaining cluster galaxies, which have not been distinguished for their accretion history, i.e. isolated or in lower mass groups. We should expect the fraction of star-forming objects to be lower in groups compared to that of cluster infall regions if pre-processing has occurred. We find the fraction of star-forming galaxies in groups does not change much with respect to cluster-centric radius, indeed suggesting already occurred pre-processing. Furthermore, the fraction of star-forming galaxies in groups is compatible with cluster cores, suggesting the effectiveness of pre-processing in quenching star formation. Hence, infalling groups appear to populate clusters with pre-processed galaxies, with significant lower values of star formation compared to isolated field objects.
 
Several are the physical processes suggested to be responsible for the fast ageing of galaxies, as a result of pre-processing. Gravitational encounters can trigger violent episodes of star formation, with a subsequent consumption and expulsion of gas \citep{park09, moore98, dekel03}. Similarly, gas can be removed via hydrodynamical interaction between the infalling galaxy and the surrounding medium, as suggested by   \citet{gunn72} and \citet{larson80} through ram-pressure stripping and starvation, respectively. 
In this work, we do not aim to disentangle the complex physical picture behind star formation quenching. Instead, we present a systematic study of a consistently selected sample of galaxy groups, showing that infalling groups are a favourable environment for pre-processing and that groups populate clusters with passive galaxies.

Group mass estimates allowed \citet{haines17} to suggest a late (below $\rm z=0.223$) but relevant contribution of $\rm\approx16\%$ to the cluster total mass due to accretion of massive groups with $\rm M_{200}>10^{13.2} \,M_{\odot}$. These findings are corroborated by a comparison with  different approaches to the analysis of simulated dark matter halos, via considering multiple cosmological models \citep{zhao09}, different simulations \citep{vandenbosch14}, or different Millennium simulation runs \citep{fakhouri10},  resulting in a broad agreement on  massive group accretion by clusters.
Morphological analysis of Subaru images is currently underway and aims to identify structural parameter trends within the population of infalling galaxies.  Additionally, high-resolution Hubble data would allow to extend the analysis beyond basic identification of discs and tidal features. A comparison with  reference cluster and field objects would allow us to obtain insights on the different evolutionary paths. In particular, regular morphology and 
high bulge-to-disk ratio would suggest that quenching occurred at high-redshift, whereas the presence of tidal features and,more generally, perturbed stellar components would highlight recent interactions in the clusters and infalling groups. A comparison with a sample of isolated groups from surveys (e.g. COSMOS, AEGIS) would also allow evolutionary biases and the influence of large scale environment on group galaxies to be quantified.

\section*{Acknowledgements}
MB, GPS, and SLM acknowledge support from the Science and Technology Facilities Council through grant number ST/N000633/1. CPH acknowledges financial support from PRIN INAF 2014.

\bibliographystyle{mnras}
\bibliography{biblio.bib}

\bsp	
\label{lastpage}

\end{document}